\pdfoutput=1 

\documentclass[aps,prb,showpacs,twocolumn,floatfix,longbibliography]{revtex4-1}

\usepackage{amsmath,amssymb,amsfonts}
\usepackage{graphicx,epsf}
\usepackage{xspace}
\usepackage{textcomp}
\usepackage{color}
\usepackage[normalem]{ulem}

\usepackage[colorlinks,bookmarks=false,citecolor=blue,linkcolor=blue,urlcolor=blue,hyperfootnotes=true]{hyperref}

\newcommand{\be}{\begin{equation}}
\newcommand{\ee}{\end{equation}}

\begin{document}

\title{Transient and Sharvin resistances of Luttinger liquids}

\author{Thomas Kloss$^1$, Joseph Weston$^2$ and Xavier Waintal$^1$}
\affiliation{$^1$Univ.\ Grenoble Alpes, CEA, INAC-Pheliqs, 38000 Grenoble, France}
\affiliation{$^2$QuTech and Kavli Institute of Nanoscience, Delft University of Technology, 2600 GA Delft, The Netherlands}

\begin{abstract}
Although the intrinsic conductance of an interacting one-dimensional system is renormalized by the electron-electron correlations,
it has been known for some time that this renormalization is washed out by the presence of the (non-interacting) electrodes to which the wire is connected.
Here, we study the transient conductance of such a wire: a finite voltage bias is suddenly applied across the wire and we measure the current before it has enough time to reach its stationary value. These calculations allow us to extract the Sharvin (contact) resistance of Luttinger and Fermi liquids. In particular, we find that a perfect junction between a Fermi liquid electrode and a Luttinger liquid electrode is characterized by a contact resistance that consists of half the quantum of conductance in series with half the intrinsic resistance of an infinite Luttinger liquid.
These results were obtained using two different methods: a dynamical Hartree-Fock approach and a self-consistent Boltzmann approach.
Although these methods are formally approximate we find a perfect match with the exact results of Luttinger/Fermi liquid theory.
\end{abstract}
\date{April 26, 2018}
\maketitle

\section{Introduction}
The theory of quantum transport describes how a quantum device is connected to the voltage sources and measuring apparatus of the macroscopic world \cite{datta97}. This description has not always been obvious, and the correct formula for the conductance of a quantum circuit was strongly debated at the beginning of quantum nanoelectronics (formerly mesoscopic physics). In its original paper \cite{landauer70}, Landauer identified the conductance with $g_{\rm I}= g_{\rm F} T/(1-T)$ where  $g_{\rm F}=e^2/h$ is the quantum of conductance and $T$ is the probability for an electron to be transmitted through the device (in one dimension, ignoring spin). It took several years for the correct expression $g = g_{\rm F} T$ ---now known as the Landauer formula--- to emerge and it was not established until the seminal experiment of Wees \textit{et al}.\ \cite{wees88} which measured the quantification of conductance in unit of $g_{\rm F}$ in a point-contact geometry. The expression $g = g_F T$ for the conductance is in itself rather spectacular: it predicts that even  in the total absence of scattering ($T=1$, perfect transmission), a quantum circuit has a finite resistance $1/g_F$, hence that some energy must be dissipated through Joule effect. The solution to this paradox comes from the concept of Sharvin ---or contact--- resistance: The expression $g = g_F T$ should be understood as 
\begin{equation}
\label{eq:sharvin}
\frac{1}{g} = \frac{1}{2g_{\rm F}} + \frac{1}{g_{\rm I}} + \frac{1}{2g_{\rm F}},
\end{equation}
i.e.\ the measured resistance $1/g$ corresponds to an intrinsic resistance of the circuit $1/g_{\rm I}$ in series with the universal Sharvin resistance $1/(2g_{\rm F})$ of each contact. More recently, the resistance entering in the RC relaxation time of a quantum capacitor (a situation with one unique electrode) has been measured explicitly \cite{Gabelli06} and corresponds to the Sharvin resistance $1/(2g_{\rm F})$ \cite{buttiker93}.
The purpose of this article is to generalize the concept of Sharvin resistance, now well understood in Fermi liquid electrodes, to interacting one-dimensional electronic systems, i.e.\  Luttinger liquids \cite{giamarchi}.

Luttinger liquids (L) can be understood formally using the bosonization formalism \cite{Haldane81,Delft98,giamarchi,Voit95}. In a simple physical picture, the excitation spectrum of a Luttinger liquid is made of plasmons \cite{Chaplik85}, ripples of the Fermi sea, that obey bosonic statistics. 
These plasmons have a linear dispersion relation $\omega = v_{\rm L} k$, where the Luttinger liquid or plasmon velocity $v_{\rm L}$ is  renormalized with respect to the Fermi velocity $v_{\rm F}$.
Likewise, the conductance $g_{\rm L}$ also gets renormalized \cite{Kane92,matveev93a} with respect to the Fermi liquid (F) result $g_{\rm F}$,
\begin{subequations}
\label{eq:vp-1d} 
\begin{align}
\frac{v_{\rm L}}{v_{\rm F}} &=  \sqrt{1 + \frac{U}{\pi v_{\rm F}}},
  \label{eq:vp-1d-1} \\
\frac{g_{\rm L}}{g_{\rm F}}  &=  \frac{v_{\rm F}}{v_{\rm L}},
  \label{eq:vp-1d-2}
\end{align}
\end{subequations}
where  $U$ characterizes the strength of the electron-electron interaction \cite{mahan,matveev93a}.
It was soon understood \cite{safi95, maslov95}, however, that the intrinsic renormalized conductance of the Luttinger liquid could not be observed directly: in presence of Fermi liquid electrodes, the effect of interaction is washed out and one recovers the Fermi liquid result $g_{\rm F}$.

In this article, we extend these results to interacting electrodes.  We study the ---transient--- response of a Luttinger liquid connected to two Fermi liquid reservoirs when a voltage bias is abruptly switched on across the wire. This technique provides a shortcut for the study of the Sharvin resistance: at short times, the system does not yet know about the presence of the Fermi liquid electrodes, so that one observes a plateau that corresponds to the conductance between two Luttinger liquids. At longer times one should recover the non-interacting Fermi liquid conductance \cite{safi95, maslov95}. The interest in the transient response of Luttinger liquids have been mostly theoretical so far \cite{Jauho94, Myohanen08,Moldoveanu10,Salvay10,Calzona15,Calzona15a,Mueller17}, with a focus on the phenomena on spin-charge separation.
However, the recent experimental progress in manipulating quantum nanoelectronic circuit at high frequencies (10 GHz and beyond) while retaining low temperatures (a few tens of a mK) now make this type of measurement feasible in the lab \cite{Dubois13,Jullien14,Grenier,Kamata14,Perfetto14,Hashisaka17}. Experiments at terahertz frequencies are also becoming possible \cite{Parkinson07,Zhong08}.

\section{Model}
We consider an infinite (quasi-) one-dimensional wire described by a  Hamiltonian of the form
$\hat H(t) = \hat H_0 + \hat H_{\rm b}(t) + \hat H_{\rm int}$.  $\hat H_0 = \sum_{<ij>,\sigma} \gamma_{ij} c^\dagger_{i\sigma} c_{j\sigma}$
contains nearest neighbor hoppings $\gamma_{ij}=-1$, while $c^\dagger_{i\sigma}$ and $c_{i\sigma}$ are the creation and destruction operators on site $i$ with spin $\sigma$.
$\hat H_{\rm b}(t) = \sum_{i\sigma} V_{\rm b}(t) \theta(i_{\rm b} -i) c^\dagger_{i\sigma} c_{i\sigma}$ corresponds to a time dependent bias voltage $V_{\rm b}(t)$ where the potential drop happens between site $i_{\rm b}$ and $i_{\rm b}+1$ ($\theta(x)$ is the Heaviside function). We will consider various shapes for the voltage pulse $V_{\rm b}(t)$ in the regime where the voltage varies quickly with respect to the propagation time through the interacting region, yet slow with respect to the bandwidth of the model (in order to avoid spurious effects associated with energy scales comparable with the Fermi energy $E_{\rm F}$). At time $t<0$ the system is at equilibrium and at zero temperature.
The interacting term is of the Hubbard form $\hat H_{\rm int} = U \sum_i s(i) (c^\dagger_{i\uparrow}c_{i\uparrow} - n_0)(c^\dagger_{i\downarrow} c_{i\downarrow} - n_0) $ where the function $s(i)=(\tanh[(i-i_L)/d] - \tanh[(i-i_R)/d])/2$ characterizes the transition between the central interacting region $[i_L,i_R]$ and the non-interacting electrodes over a width $d$, while $n_0$ is the equilibrium density.
A sketch of the system is shown in Fig.\ \ref{fig:vp-1d-sim}. We consider two (formally) approximate methods to study this problem: a time dependent Hartree-Fock approach which is equivalent to the Random Phase Approximation, hence already known to capture the salient features of the one dimensional plasmons \cite{giamarchi} and a much simpler self-consistent Boltzmann approach that provides analytical expressions
in a number of situations.

\section{Time dependent Hartree-Fock approach} 
The first method is a time dependent
mean-field approach where $\hat H_{\rm int}$ is replaced
by its mean-field value $\hat H_{\rm HF}$ at time $t$.
 The problem becomes effectively diagonal in spin, so that from now on we ignore the spin degree of freedom and consider the spinless problem (the spinfull conductance can be recovered by multiplying our results by a factor 2). Note that this time dependent Hartree-Foch approach is very different from its static counterpart; in particular it captures screening effects at the Random Phase Approximation level \cite{negele_orland}. The interaction term takes the form $\hat H_{\rm HF} = U \sum_i s(i) c^\dagger_{i} c_{i} \left[\langle c^\dagger_{i}(t) c_{i}(t)\rangle - n_0 \right]$, where the notation $c_{i}(t)$,  $c^\dagger_{i}(t)$ refers to the Heisenberg representation.  We solve this problem using the method developed in Refs.\ [\onlinecite{gaury14,weston15,weston16}] within the Keldysh formalism which is implemented as a top layer of the Kwant package \cite{groth14}. It amounts to solving a set of one-particle Schr\"odinger equations
\begin{equation}
	i \partial_t \psi_{\alpha E}(i,t) = \sum_j H_{ij}(t)\psi_{\alpha E}(j,t),
	\label{eq:se_time_dep}
\end{equation}
with $H_{ij} = \gamma_{ij} + V_{\rm b}(t) \theta(i_{\rm b} -i) + U s(i) (n(i,t)-n_0)$ and the initial condition
$\psi_{\alpha E}(i,t=0) = \psi_{\alpha E}^{\rm st}(i)$ where $\psi_{\alpha E}^{\rm st}(i)$ is a scattering state $\alpha$ at energy $E$ of the time-independent Hamiltonian $\hat H_0$ at $t=0$. The local density of electrons $n(i,t)$ is given by,
\begin{equation}
n(i, t) = \sum_\alpha \int \frac{dE}{2\pi} f_0(E) |\psi_{\alpha E}(i,t)|^2
\label{eq:kwant_normalization}
\end{equation}
where $f_0$ is the Fermi function.
The new twist with respect to Refs.\ [\onlinecite{gaury14,weston15,weston16}] is the presence of the self-consistent term $U s(i) (n(i,t)-n_0)$ which transforms the set of initially independent Schr\"odinger equations into coupled ones. Since up to 1000 different energy values are necessary to discretize the integral of Eq.\ (\ref{eq:kwant_normalization}), such a set of coupled partial integro-differential equations is in general intractable. We leverage the
fact that the dynamics of Eq.\ (\ref{eq:kwant_normalization}) is much slower than the one of Eq.\ (\ref{eq:se_time_dep}) to develop a doubly adaptative timestepping scheme: we construct a linear extrapolation of $n(i,t)$ which is used by a Runge-Kutta method with
Dormand and Prince stepsize control \cite{hairer93} to integrate the one-particle Schr\"odinger equations (\ref{eq:se_time_dep}). At a second larger (and also adaptive) time step, the integral Eq.\ (\ref{eq:kwant_normalization}) is recalculated and the extrapolation of $n(i,t)$ is updated. In all the simulations below, the overall error in the calculated observables is of the order $10^{-4}$.

\begin{figure}[tb]
	\centering
	\includegraphics[width=80mm]{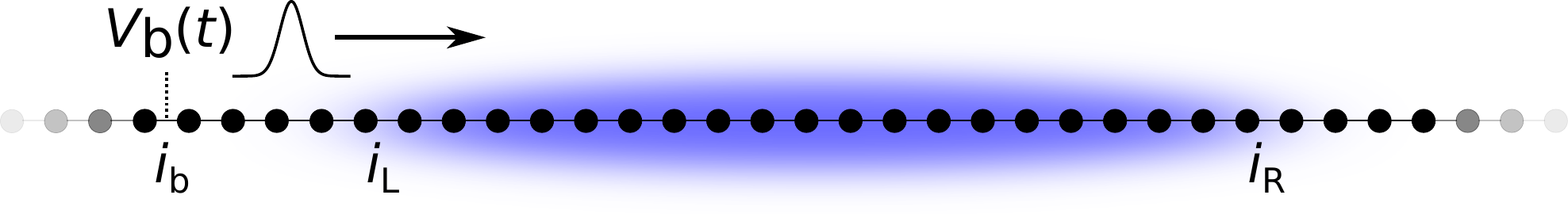}
	\vspace{2mm}
	{
	\includegraphics[width=44mm]{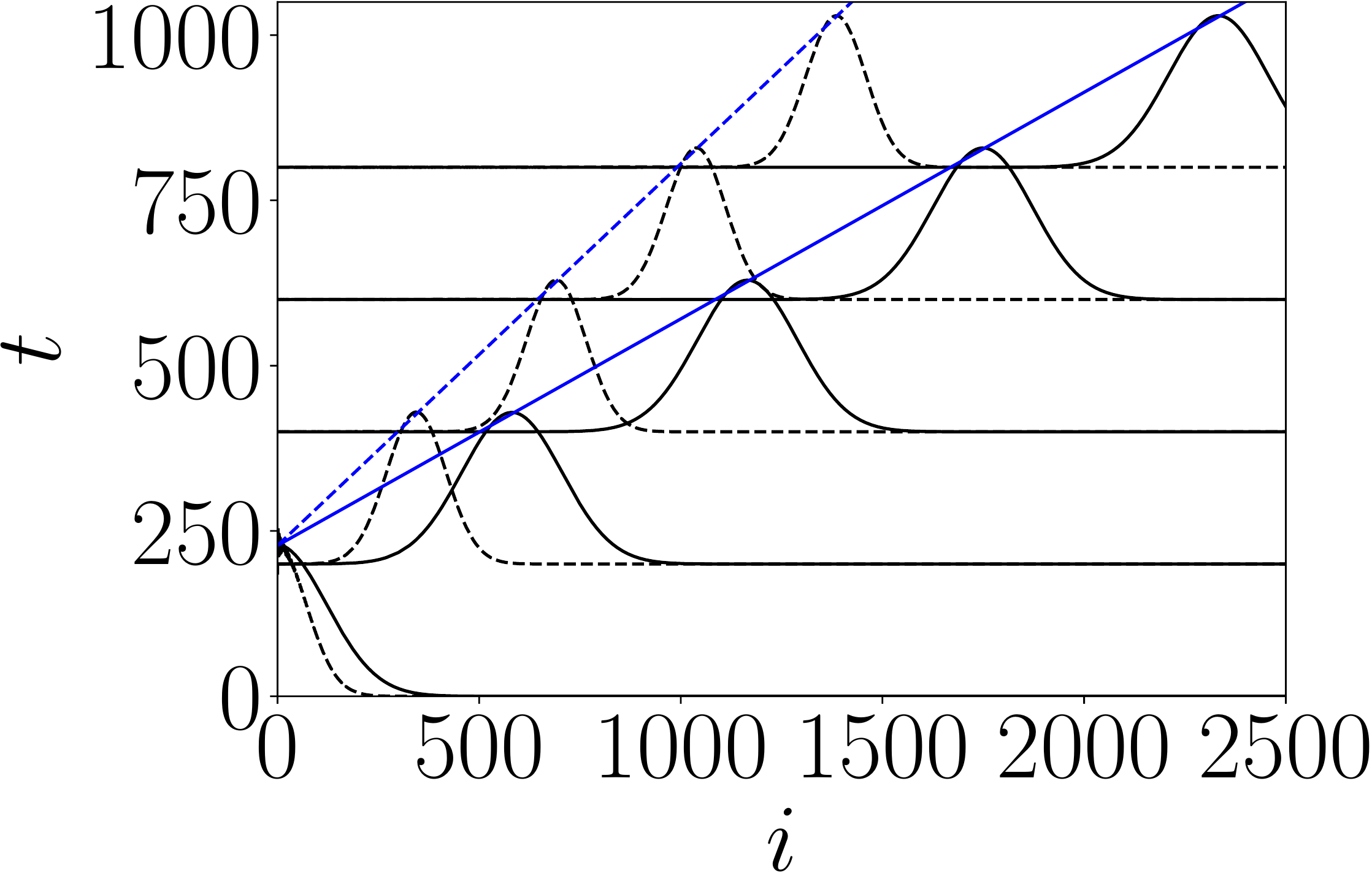}
	\hspace{-1ex}
	\includegraphics[width=37mm]{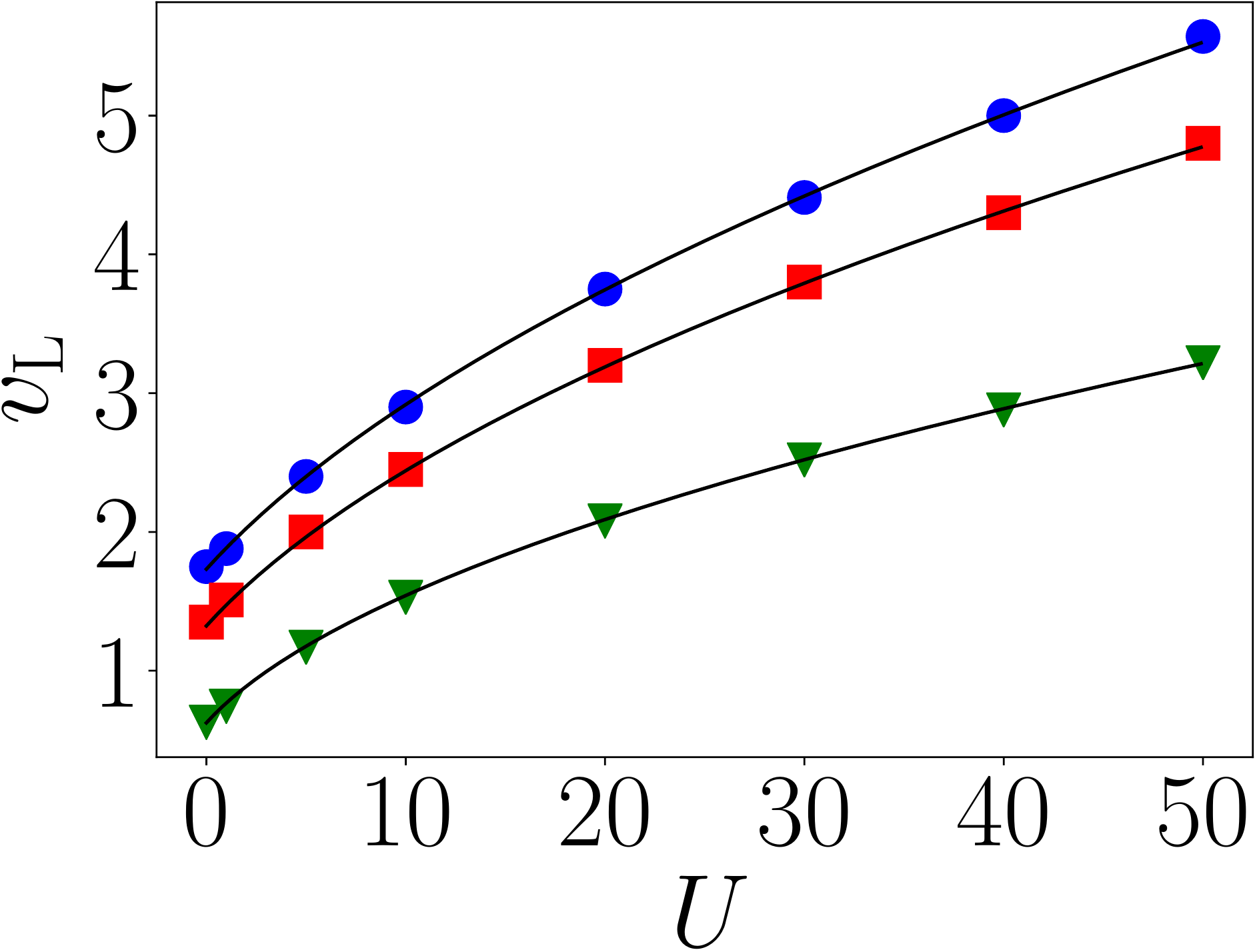}
	}
	\vspace{-4.0mm}
  \caption{%
(Color online)
Top panel: sketch of the system under consideration including the position of the voltage drop (electrode/wire interface) $i_{\rm b}$ and interacting blue region. Transport simulation results for the Plasmon velocity $v_{\rm P}$ in one-dimensional quantum wires.
Left panel: electronic density $n(i,t)$ (in arbitrary units) as a function of site  $i$ for different values of time $t$ (as indicated on the $y$-axis) after injection of a Gaussian voltage pulse of width $\tau = 100$. The interacting region is $|i_R-i_L|=1000$ sites wide. Solid lines: $U = 10$, dashed lines: $U = 0$ (no interaction). The blue lines are linear fits from which we extract the velocity.
Right panel: extracted Plasmon velocity $v_{\rm L}$ vs.\ interaction strength $U$ for three different Fermi energies
$E_{\rm F} = -1, \ -1.5$ and $-1.9$. Symbols are from the simulations and lines correspond to Eq.\ (\ref{eq:vp-1d-1}).
 }
  \label{fig:vp-1d-sim}
\end{figure}

Figure \ref{fig:vp-1d-sim} shows a typical simulation for the propagation of a voltage pulse through the (blue) interacting region.
The left panel shows two simulated charge densities $n(i,t)$, one with interactions (solid line) and one without interactions (dashed line). The injected Gaussian pulse $V_{\rm b}(t)= V_{\rm b} e^{-(t-t_0)^2/2\tau^2}$ propagates faster in the interacting case. The right panel of Fig.\ \ref{fig:vp-1d-sim} shows the dependence of the plasmon velocity $v_{\rm L}$ on the interaction strength $U$ for three different Fermi energies. We find a perfect match between the numerical data (symbols, extracted from a linear fit of the numerical simulations as shown in the blue lines of the left panel) and the Luttinger-liquid prediction Eq.\ (\ref{eq:vp-1d-1}) (solid lines). The fact that the dispersion relation of the Luttinger liquid is perfectly reproduced by a mean-field approach is in itself nontrivial \cite{giamarchi}.

We note that Equation (\ref{eq:vp-1d}) together with the Luttinger liquid approach of the one dimensional Hubbard model is only valid for moderate interactions. 
When $U$ becomes of the order of the bandwidth (e.g. $U\sim 2$ in our case), the plasmon physical picture ceases to be relevant and one needs to resort to other approaches such as the Bethe ansatz\cite{Schultz91, essler_book}. Likewise, the results presented here do not apply in the strongly interacting limit. However, since we are chiefly interested in the out-of-
equilibrium behavior of Luttinger liquids, losing the relation to
the underlying microscopic model is not necessarily an issue.

\subsection{Transient conductances} 
\begin{figure}[tbh]
	\centering
	{
	\includegraphics[width=80mm]{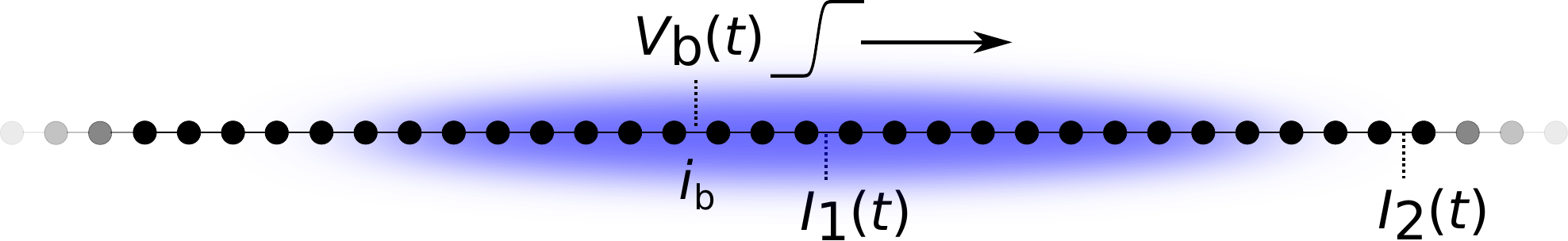}
	\vspace{-0.5ex}
	}
	{\includegraphics[width=70mm]{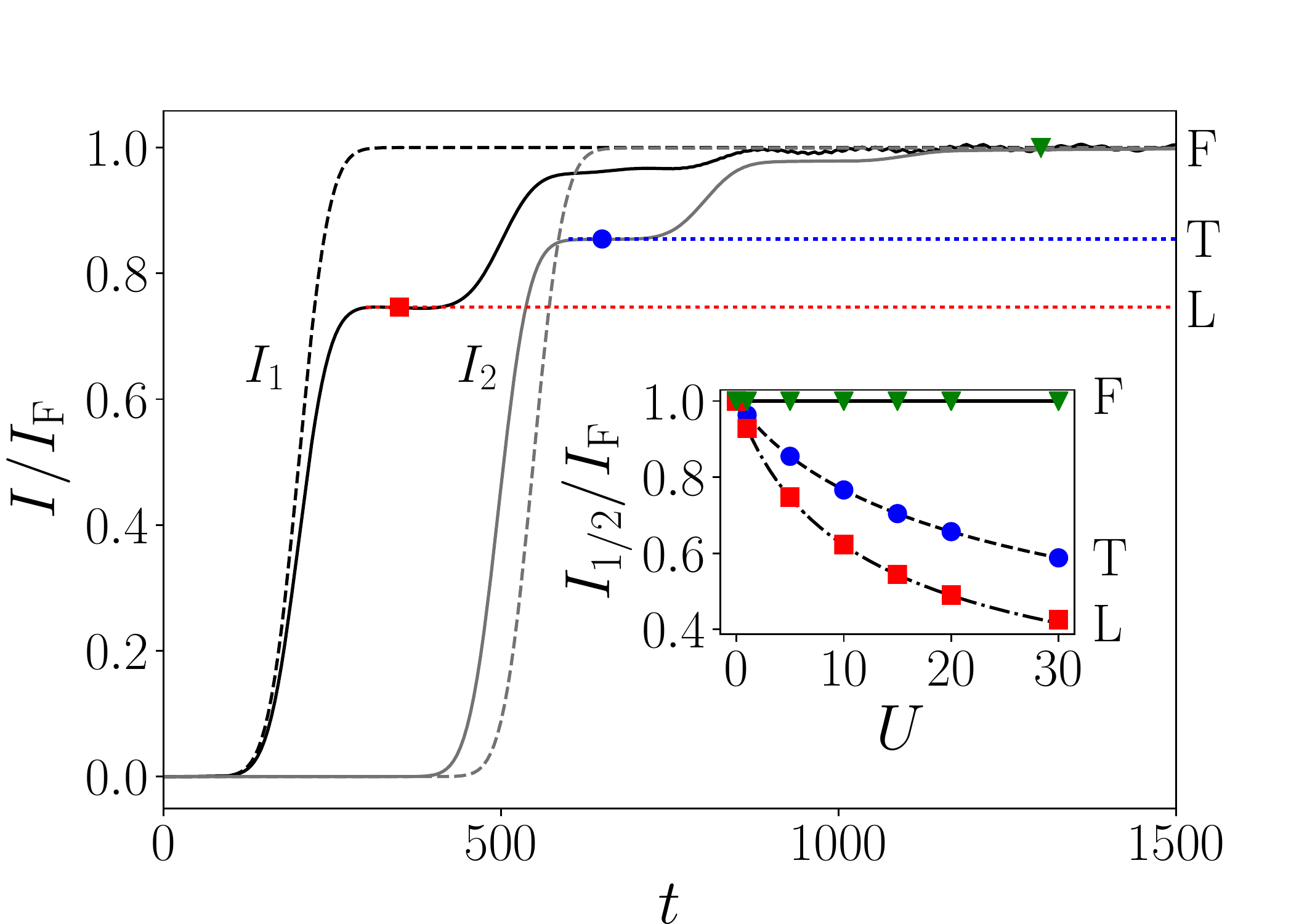}
	\vspace{-3mm}
	}
  \caption{%
(Color online)
Numerical simulation of the current $I$ as a function of time $t$ in a one-dimensional quantum wire in presence (solid lines, $U=5$) and absence (dashed lines, $U=0$) of interactions. The electrical potential $V_{\rm b}$ is switched on smoothly at time $t_0 = 200$ and stays constant after. The subscript on $I_{1/2}$ indicates the position where the current is actually measured, as shown in the sketch.
Inset: transient conductances as a function of $U$, see text (symbols) and theoretical predictions (dashed line: transient  Eq.\ (\ref{eq:1d-transient-conductivit}), dashed dotted: Luttinger liquid Eq.\ (\ref{eq:vp-1d-2}) and solid line: Fermi liquid).
 }
  \label{fig:current-1d-center-1}
\end{figure}

\begin{figure}[tbh]
	\centering
	{
	\includegraphics[width=80mm]{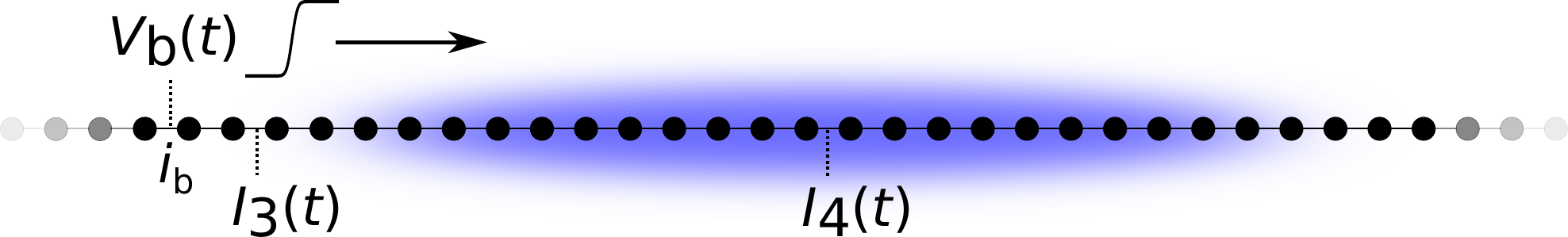}
	\vspace{-0.0ex}
	}
	{\includegraphics[width=70mm]{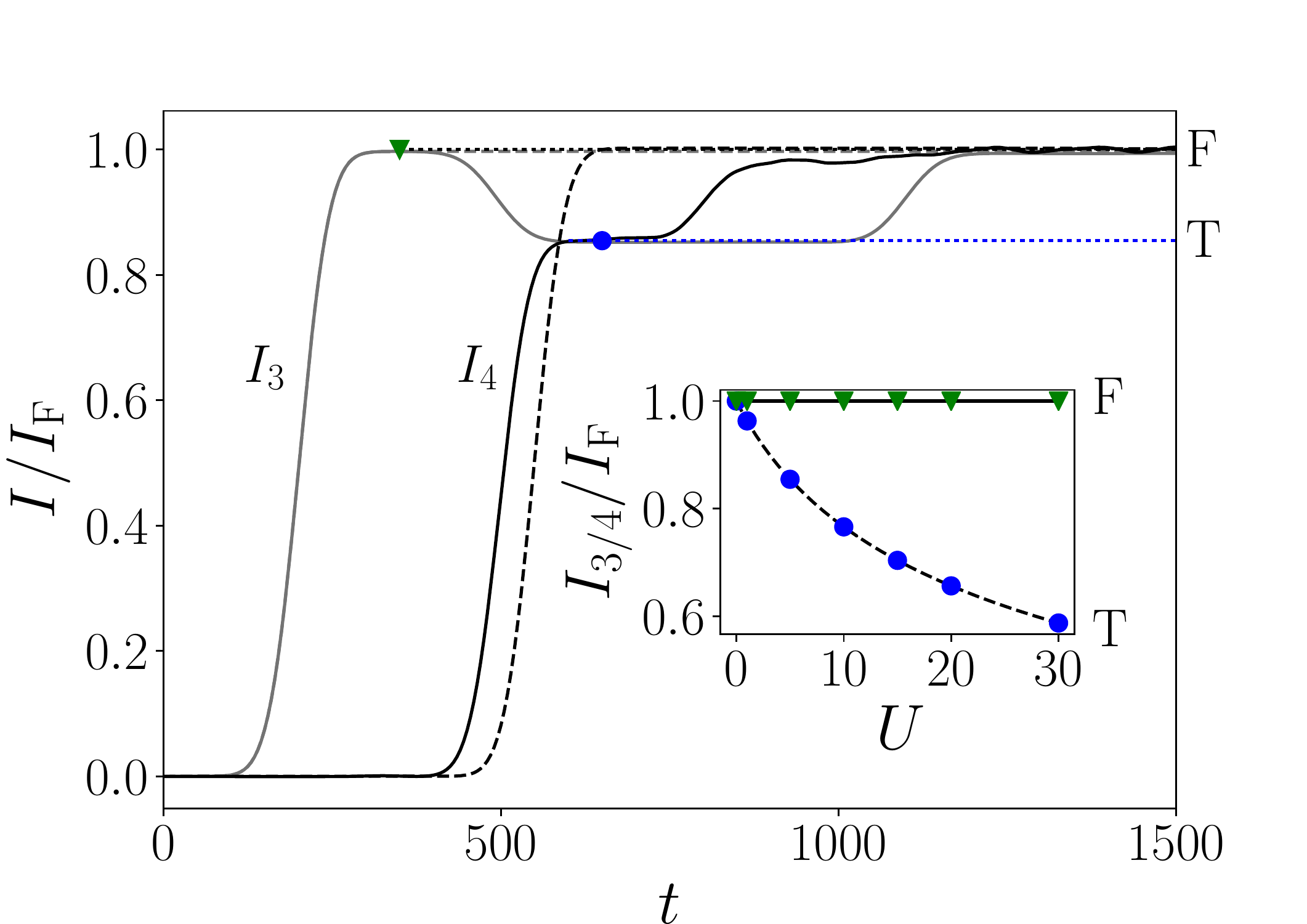}
	\vspace{-3mm}
	}
  \caption{%
(Color online) Same as Fig.\ \ref{fig:current-1d-center-1} but the voltage drop happens at the electrode wire interface.
 }
  \label{fig:current-1d-center-2}
\end{figure}

We now turn to the transient conductances of our system:
 at $t=0$ we abruptly raise the voltage bias across the wire from zero to $V_{\rm b}$ and measure the current $I$ flowing through the system. The results are presented in Fig.\ \ref{fig:current-1d-center-1} and Figure \ref{fig:current-1d-center-2} where 
 solid (resp. dashed) lines mark the result with (without) interaction.
The solid line of Fig.\ \ref{fig:current-1d-center-1} shows the current $I_1$ inside the interacting region as a function of time after a voltage drop in the interacting region.
We observe a well developed plateau until the pulse reaches the boundary of the interacting region, followed by a second, stationary, plateau. The duration of the first plateau is simply given by the propagation time $|i_R-i_L|/v_{\rm P}$ in the interacting region. 
We define the (dimensionless) transient conductance $g$ as $I/I_{\rm F}$, where the current is evaluated on a plateau and $I_{\rm F} = V_{\rm b}/2\pi$ is the non-interacting value. Note that since the system is not stationary, the transient conductance varies inside the system, in particular depending on the measurement being performed inside the wire versus inside the electrodes.
The value of the transient conductance is illuminating: the short time value is given by the Luttinger liquid theory Eq.\ (\ref{eq:vp-1d-2}) while the long
time (stationary) value is given by the Fermi liquid value $I_1/I_{\rm F} = 1$, see the inset of
Fig.\ \ref{fig:current-1d-center-1}. It is by itself rather remarkable that our mean-field approach captures these two values exactly.
The  transient conductance $I_2/I_{\rm F}$ measured inside the electrode is however very different: it is given by a new formula 
\begin{equation}
\frac{1}{g_{\rm T}} = \frac{1}{2} \left[ \frac{1}{g_{\rm L}} + \frac{1}{g_{\rm F}} \right],
\label{eq:1d-transient-conductivit}
\end{equation}
which corresponds to half a quantum of resistance in series with half the Luttinger liquid intrinsic resistance.
This is our first hint that $1/(2g_L)$ should be understood as the Sharvin resistance of a Luttinger liquid.
Figure \ref{fig:current-1d-center-2} is very similar to Fig.\ \ref{fig:current-1d-center-1} but the voltage drop happens at the  electrode-wire interface. There, only
the T transient conductance is observed. In reality the position of the drop is determined by the interplay between the electrostatic capacitance and the quantum capacitance of the circuit (see [\onlinecite{gaury14}] for a discussion). In practice,
the drop is most likely to happen at the electrode, making Fig.\ \ref{fig:current-1d-center-2} the experimentally most relevant case.

\section{Self consistent Boltzmann approach} To proceed, we introduce a second, even simpler approach, in the spirit of
the classical theory of surface plasmons \cite{Chaplik85}. We introduce the semi-classical probability distribution $f(x,k,t)$
for an electron to be at position $x$ (we use $x$ instead of $i$ as we will be working in the continuum limit) and momentum $k$ at time $t$. In our ballistic sample, $f(x,k,t)$  satisfies a collisionless Boltzmann equation \cite{kamenev_book},
\begin{equation}
\partial_t f =  - v_{k} \partial_x f - F(x,t) \partial_k f,
\label{eq:boltzmann}
\end{equation}
where $v_k = dE/dk$ is the non-interacting velocity corresponding to the dispersion relation $E(k)= -2\cos(k)$ of $\hat H_0$.
The classical force $F(x,t)$ arises from the electric field of the bias potential $V_{\rm b}(x,t)$ and from the field generated by the
pulse itself through Coulomb repulsion $F(x,t) = \partial_x V_{\rm b}(x,t) + U \partial_x [s(x) n(x,t)]$. The density $n(x,t)$
is obtained through
\begin{equation}
n(x,t) = \int_{-\pi}^\pi \frac{dk}{2\pi} f(x,k,t)
\end{equation}
and we arrive at a close set of equations. 

\subsection{General solution}
Interestingly, the above self-consistent Boltzmann equation reproduces exactly the plasmon
velocity and conductance of the Luttinger liquid Eq.\ (\ref{eq:vp-1d-1}) and (\ref{eq:vp-1d-2}) as well as the transient regime Eq.\ (\ref{eq:1d-transient-conductivit}) that will be discussed below. The low energy spectrum of the Boltzmann equation can be obtained by expanding $f(x,k,t) =
f_0[E(k)] + f_1(x,k,t)$ to first order in $f_1$ around the Fermi function $f_0$. After integration over $k$, we introduce the density of right $n_+(x,t)= \int_0^\pi \frac{d k}{2 \pi} f_1$ and left
$n_-(x,t) = \int_{-\pi}^0 \frac{d k}{2 \pi} f_1$   movers and arrive at
\begin{equation}
\partial_t n_{\pm} \pm  \partial_x \left[ v_{\rm F} n_{\pm} + \frac{U}{2\pi} s(x) n_1 - \frac{V_{\rm b}}{2\pi}\right] = 0 ,
\label{eq:boltzmann_lr}
\end{equation}
with $n_1 = n_+ + n_-$. Equation (\ref{eq:boltzmann_lr}) has the form of a Liouville equation, hence describes ballistic propagation of different modes. In particular for a constant interaction $s(x)=1$, it can be solved exactly and its general solution reads (in terms of the initial condition at $t=0$),
\begin{align}
&n_{\pm}(x,t) =  \frac{1}{4v_{\rm F} v_{\rm L}} \sum_{\eta =\pm 1} \eta \biggl[  (v_{\rm F} + \eta v_{\rm L})^2 n_\pm (x \mp \eta v_{\rm L} t, t=0) \nonumber \\
& - (v_{\rm F}^2 -  v_{\rm L}^2) \ n_\mp (x \mp \eta v_{\rm L} t, t=0)  \nonumber \\
&+  2 v_{\rm F} \left(v_{\rm F} \pm \eta v_{\rm L} \right) \int_0^t \frac{dt'}{2\pi} \frac{\partial}{\partial x} V_{\rm b}(x -\eta v_{\rm L} (t-t'),t') \biggr]
\label{eq:sol_inh}
 \end{align}
and indeed propagates with the plasmon velocity $v_{\rm L}$ given by Eq.\ (\ref{eq:vp-1d-1}). This theory can be generalized
to the multi-channel situation, and one recovers the same theory as what can be derived from the bosonization approach \cite{matveev93,matveev93a}. The multi-channel theory has recently been verified experimentally in a wave guide geometry \cite{roussely17}
and is consistent with experiments in the integer quantum Hall regime \cite{Hashisaka17}.

\subsection{Derivation of the transient conductances from the Boltzmann approach} 
Let us now derive the values of the transient conductances
from the Boltzmann approach. In a transient regime where the densities $n_+(x)$ and $n_-(x)$ are {\it locally } stationary (they may still propagate in a remote region), Eq.\ (\ref{eq:boltzmann}) admit two constants of motion, the current $I$ and the renormalized density $N$,
\begin{subequations}
\begin{align}
I &= v_{\rm F} [n_+(x) - n_-(x) ]  \ \ \forall \, x
\label{current}
\\
N &= \left(1 + \frac{U s(x)}{\pi}\right) [n_+(x) + n_-(x) ] \ \  \forall \, x
\label{eq:N}
\end{align}
\end{subequations}
Let us now consider the general situation of a junction between one Luttinger liquid on the left [characterized by an interaction parameter
$U_1$ and the corresponding conductance $g_{\rm L1}$, see Eq.\ (\ref{eq:vp-1d})] in contact with a second Luttinger liquid on the right ($U_2$ and $g_{\rm L2}$).
At $t=0$, $n_+(x,t = 0) = n_-(x, t = 0) = 0$ and we suddenly raise a sharp potential step $V_{\rm b}(x) = V_{\rm b} \theta(x_1-x)\theta(t)$
with $x_1$ deep in the left electrode. Deep in the left and right electrode (where $s(x)$ = constant), we can use Eq.\ (\ref{eq:sol_inh})
and obtain the structure of the solution; in a second step, these solutions are matched using Eqs.\ (\ref{current}) and (\ref{eq:N}). After some algebra, we arrive at  $I = g V_{\rm b}$ with,
\begin{equation}
\frac{1}{g} = \frac{1}{2g_{\rm L1}} + \frac{1}{2g_{\rm L2}},
\label{eq:U1U2}
\end{equation}
which is our chief analytical result. Eq.\ (\ref{eq:U1U2}) calls for a number of comments. 

(i) First, it is valid for an arbitrary function $s(x)$, i.e.\ the conductance only depends on the nature of the left and right electrodes and {\it not } on the intermediate region. 

(ii) In particular, in the case of a Luttinger liquid sandwiched in between non-interacting electrodes  [$s(0)=1$, $s(\pm\infty)=0$], one recovers the established fact that the electron-electron interaction does not renormalize the conductance \cite{safi95, maslov95}.

(iii) For an infinitely long and homogeneously interacting wire [$s(x)=1$], we recover the Luttinger liquid result $g=g_{\rm L}$.

(iv) For an non-interacting electrode in contact with an interacting one, Eq.\ (\ref{eq:U1U2}) reduces to Eq.\ (\ref{eq:1d-transient-conductivit}).

\section{Discussion and Outlook}
Equation (\ref{eq:U1U2}) is strikingly similar to Eq.\ (\ref{eq:sharvin}) that defines the Sharvin resistance of a non-interacting electrode.
In fact, Eq.\ (\ref{eq:U1U2}) can be interpreted as the generalization of the concept of Sharvin resistance to Luttinger-liquid electrodes, attributing the contact resistance $1/(2g_{\rm L})$ to each electrode. 
We have studied the perfectly transmitting situation (e.g.\ the absence of impurities that could lead to scattering) that corresponds to the absence of intrinsic resistance. 
The generalization to the presence of impurities leads to power law behaviors in the $I(V_{\rm b})$ which is beyond the scope of the Boltzmann approach \cite{Voit95,Li91}. 
The concept of Sharvin resistance, however, can be used beyond the present situation. Indeed, what defines an electrode in practice depends on where the energy relaxation takes place; an electrode is essentially any system whose size is comparable to its energy relaxation length. 
Hence, we predict that the renormalization of the conductance Eq.\ (\ref{eq:U1U2}) should be amenable to observation in d.c.\ experiments involving interacting wires of sufficient length. For instance, a geometry close to the one used in [\onlinecite{roussely17}] should allow for the determination of the interacting Sharvin resistance with  $U\sim d/a_B$ ($d$ distance to the gate, $a_B$ effective Bohr radius). The above results might also be measured directly in fast transient experiments, a regime whose experimental study has only recently began. 
Our identification of the interacting Sharvin resistance also implies that the relaxation time of a quantum capacitor connected to a Luttinger liquid is controlled by $1/(2g_{\rm L})$ \cite{Hamamoto10}. This last experiment could be performed, for instance, by extending the measurements of Ref.\ [\onlinecite{Gabelli06}] to the fractional quantum Hall regime.

Finally, let us note that besides the presented context of nanoelectronics,
our approach could also prove useful in other areas such as
cold atoms, see e.g.\ [\onlinecite{Uehlinger13}] and [\onlinecite{Langen15}], where Feshbach resonances allows to tune the interactions over a wide range.

\section*{ACKNOWLEDGMENTS} 
Interesting discussions with Chris Bauerle, Christian Glattli, Patrice Roche and Fabien Portier are gratefully acknowledged. This work is funded by the French ANR QTERA, ANR FullQuantum and the US Office of Naval Research.

\bibliography{meso}
\bibliographystyle{apsrev4-1_own}

\end{document}